\pdfoutput=1
\documentclass[a4paper,fleqn,usenatbib,useAMS]{mnras}

\usepackage{newtxtext,newtxmath}

\usepackage{graphicx}	
\usepackage{amsmath}	
\usepackage{amssymb}	
\usepackage{multicol}        
\usepackage{bm}		
\usepackage{pdflscape}	
\usepackage[T1]{fontenc}
\usepackage{ae,aecompl}
\usepackage{newtxtext,newtxmath}
\usepackage{float}
\usepackage{epsfig}

\usepackage[caption =false]{subfig}
\usepackage{float}

\title[Lyman-$\alpha$ Emitters in the Reionization-Era]{Spectroscopic Constraints on UV Metal Line Emission at $z\simeq 6-9$: 
The Nature of Ly$\alpha$ Emitting Galaxies in the Reionization-Era}

\author[Mainali et al.]{Ramesh Mainali,$^{1}$\thanks{E-mail:rmainali@email.arizona.edu}
Adi Zitrin$^{2}$,
Daniel P. Stark$^{1}$,
Richard S. Ellis$^{3}$,
Johan Richard$^{4}$, \newauthor
Mengtao Tang$^{1}$,
Nicolas Laporte$^{3}$,
Pascal Oesch$^{5}$,
\& Ian McGreer$^{1}$
\\
$^{1}$Steward Observatory, University of Arizona, 933 N Cherry Ave, Tucson, AZ,USA\\
$^{2}$Physics  Department,  Ben-Gurion  University  of  the  Negev,
P.O. Box 653, Be'er-Sheva, 84105, Israel\\
$^{3}$Department of Physics and Astronomy, University College London, Gower Street, London WC1E 6BT, UK\\
$^{4}$ Univ Lyon, Univ Lyon1, Ens de Lyon, CNRS, Centre de Recherche Astrophysique de Lyon UMR5574, F-69230, Saint-Genis-Laval, France\\
$^{5}$Geneva Observatory, University of Geneva, Ch. des Maillettes 51, 1290 Versoix, Switzerland
}

\pubyear{2018}

\begin{document}
\label{firstpage}
\pagerange{\pageref{firstpage}--\pageref{lastpage}}
\maketitle

\begin{abstract}
Recent studies have revealed intense UV metal emission lines  in  a modest sample of $z>7$ 
Lyman-$\alpha$ emitters, indicating a hard ionizing spectrum is present. If such high ionization features 
are shown to be common, it may indicate that extreme radiation fields play a role in regulating  
the visibility of Ly$\alpha$ in the reionization era.   Here we present deep  near-infrared 
spectra of seven galaxies with Ly$\alpha$ emission at $5.4<z<8.7$ (including a newly-confirmed 
lensed galaxy at $z_{\rm{Ly\alpha}}=6.031$) and three  bright $z\simeq 7$ photometric targets.   
In nine sources we do not detect UV metal lines.   However in the $z_{\rm{Ly\alpha}}=8.683$ galaxy 
EGSY8p7, we detect a 4.6$\sigma$ emission line in the narrow spectral window expected for NV$\lambda$1243.  
The feature is unresolved (FWHM$<$90 km s$^{-1}$) and is likely nebular in origin.   A deep H-band spectrum 
of EGSY8p7 reveals non-detections of CIV, He II, and OIII].  The presence of NV requires a substantial flux of 
photons above 77 eV, pointing to a hard ionizing spectrum powered by an AGN or fast radiative shocks.   
Regardless of its origin, the intense radiation field 
of EGSY8p7 may aid the transmission of Ly$\alpha$ through what is likely a partially neutral IGM. 
With this new detection, five of thirteen known Ly$\alpha$ emitters at $z>7$ have now been shown to have   
intense UV line emission, suggesting that
extreme radiation fields are  commonplace among the Ly$\alpha$ population.  
Future observations with  {\it JWST} will eventually clarify the origin of these 
features and explain their role in the visibility of Ly$\alpha$ in the reionization era.
\end{abstract}

\begin{keywords}
galaxies -- high-z --reionization
\end{keywords}



\section{Introduction}
The reionization of the intergalactic medium (IGM) is a pivotal milestone in the early universe, marking 
the point at which almost every baryon has been impacted by structure formation.  Over the past two decades, 
substantial efforts have been devoted to improving our understanding of the process.  
While our  knowledge remains very limited, a  picture describing the most basic details of reionization has emerged.  
Measurement of the optical depth to electron scattering of the cosmic microwave background
photons \citep{Planck2016} and deep spectra of high redshift quasars (e.g., \citealt{McGreer2015}) 
indicate that the process is underway by $z\simeq 9$ and is completed at or before $z\simeq 6$. 
Deep imaging surveys with the {\it Hubble Space Telescope} (HST) have enabled the identification of large numbers 
of color-selected galaxies over this redshift range (e.g., \citealt{Mclure2013,Bouwens2015, Finkelstein2015,Atek2015,Livermore2017,Ishigaki2018}; for a review 
see \citealt{Stark2016}), revealing that faint star forming galaxies may provide the dominant source of photons 
required for ionizing intergalactic hydrogen (e.g., \citealt{Robertson2015, Bouwens2015b, Stanway2016}).   

Observational efforts are now focused on mapping the evolution of the IGM at $7<z<8$, when the bulk of 
reionization is thought to occur.  Most of what we currently know about the IGM 
 at these redshifts comes from measurements of Ly$\alpha$ emission from 
star forming galaxies.  As the IGM becomes partially neutral,  the damping wing 
absorption of intergalactic hydrogen begins to attenuate Ly$\alpha$ 
emission, reducing the visibility of Ly$\alpha$ emission in star forming galaxies.  
The decline of the Lyman-$\alpha$ emitter (LAE) population can be probed by the luminosity function of 
narrowband-selected Ly$\alpha$ emitters (e.g., \citealt{Malhotra2004, Kashikawa2006, 
Ouchi2010, Hu2010, Konno2018, Ota2017}) or the evolution of the Ly$\alpha$ equivalent 
width (EW) distribution in continuum-selected galaxies (e.g.,  \citealt{Stark2010,Fontana2010,Ono2012,
Schenker2014, Pentericci2014, Tilvi2014,Caruana2014, Schmidt2016}).  
Both tests demonstrate that the visibility of Ly$\alpha$ emitting galaxies drops significantly in the 170 Myr  
between $z\simeq 6$ and $z\simeq 7$.  To explain the magnitude of this drop, it is thought that the IGM 
must become significantly neutral  ($x_{\rm{HI}}=0.4-0.6$) by $z\simeq 7$ (e.g., \citealt{Dijkstra2014, 
Mesinger2015, Mason2017}).   While such results were initially met 
with skepticism, the first handful of quasars at $z>7$  have suggested similarly large neutral hydrogen fractions 
(e.g., \citealt{Mortlock2011, Greig2017, Banados2018}; c.f.\citealt{ Bosman2015}).   

Progress in identifying Ly$\alpha$ at $z>7$ has recently begun to ramp up, revealing 
a  more complex picture than described above.  Much of the  success has come from a 
photometric selection of four $z>7$ galaxies in the CANDELS fields (\citealt{RobertsBorsani2016}; hereafter RB16).  Each system is very bright ($H\simeq 25$), massive ($\sim10^{10}$ M$_\odot$ in stars), and red  
in {\it Spitzer}/IRAC [3.6]-[4.5] color, indicating the presence of very strong [OIII]+H$\beta$ emission.
Follow-up Keck spectroscopy has shown that Ly$\alpha$ emission is present in all four galaxies 
(\citealt{Oesch2015,Zitrin2015, Stark2017}; RB16), including record-breaking detections at $z=7.73$ \citep{Oesch2015} and $z=8.68$ 
\citep{Zitrin2015}.   The large Ly$\alpha$ emitter fraction in the RB16 sample is  greater than the success rates found in 
other galaxy samples at similar redshifts.  How Ly$\alpha$ emission  can be  transmitted so effectively  from  these four 
galaxies while being so strongly attenuated from most other systems at $z>7$ is still not clear.

The conventional explanation is that because of their high mass, the RB16 galaxies trace a clustered population within overdense 
regions which ionize their surroundings early.  As a result, they should be situated in large ionized patches 
of the IGM, increasing the transmission of Ly$\alpha$ \citep{Mason2018}.  The visibility of Ly$\alpha$ in high mass galaxies is 
also favored by the line profile of Ly$\alpha$ that emerges following transfer through the circumgalactic medium.  
With both a larger line width and velocity offset  than in lower mass galaxies \citep{Erb2014,Stark2017}, a 
larger fraction of Ly$\alpha$ in high mass galaxies will be redshifted far enough away from line center to escape the 
strong damping wing absorption of the neutral IGM \citep{Stark2017,Mason2017}.  
Nevertheless surveys for Ly$\alpha$ in other massive star forming galaxies at $z>7$ 
have been met with a much lower success rate than found in the RB16 sample (e.g. \citealt{Schenker2014, Bian2015,
Furusawa2016}).   

An alternative explanation that is gaining traction posits a link between Ly$\alpha$ visibility 
at $z>7$ and the intensity of the galaxy ionizing spectrum.  The large 
[OIII]+H$\beta$ emission of both the RB16 galaxies (and several other  $z>7$ Ly$\alpha$ emitters) signals 
an extreme radiation field that might indicate larger than average Ly$\alpha$ production rates, boosting the 
likelihood of detection.   
The connection between Ly$\alpha$ visibility and the hardness of the ionizing spectrum has been 
strengthened with the discovery  of intense UV metal line emission (CIII], NV, He II) in two of the RB16 galaxies 
\citep{Stark2017,Laporte2017} and two additional color-selected Ly$\alpha$ emitters at $z>7$ 
\citep{Stark2015b, Tilvi2016}.  Such intense high ionization features are extraordinarily 
rare at lower redshift, likely requiring the presence of an AGN or metal poor stellar population 
\citep{Feltre2016, Gutkin2016, Jaskot2016, Nakajima2017}.  
These results indicate that variations in radiation field may play an important role in regulating 
Ly$\alpha$ visibility at $z>7$; failure to account for such variations would lead to an incorrect 
IGM ionization state.  

Here we present new constraints on the presence of high ionization features in Ly$\alpha$ emitters at $z>7$, 
including the first investigation of UV metal lines in the $z=8.68$ galaxy EGSY8p7.  We aim to determine 
whether extreme radiation fields are present in all Ly$\alpha$ emitters in the reionization-era and 
to better characterize the far-UV spectra of those $z>7$ systems with existing metal line detections.  
We detail the spectroscopic observations in \S2 and describe the spectra of individual sources in \S3.  In \S4, we discuss 
implications for the nature of Ly$\alpha$ emitters and review what is known about the distribution of 
UV metal line equivalent widths at high redshift.

We adopt a $\Lambda$-dominated, flat Universe with $\Omega_{\Lambda}=0.7$, $\Omega_{M}=0.3$ and
$\rm{H_{0}}=70\,\rm{h_{70}}~{\rm km\,s}^{-1}\,{\rm Mpc}^{-1}$. All
magnitudes in this paper are quoted in the AB magnitude system \citep{Oke1983}, and all equivalent widths are
 quoted in the rest-frame.

\section{Observations} \label{sec:observations}

\begin{table*}
\begin{tabular}{lccccccccc}
\hline 
Source  & Redshift  & RA & DEC & Date of Observations  &  Filters & t$_{\rm exp}$ (hr) &  UV lines  targeted  & Ref \\  \hline  \hline
&&&&Spectroscopic Targets&&&& \\  \hline  \hline
EGSY8p7 &  8.683   &  14:20:08.50 &  +52:53:26.6  & 29-30 Apr 2016  & H & 5.93 &  CIV, He II, OIII]  & [1],[2]  \\
EGS-zs8-1 &  7.730  & 14:20:34.89 & +53:00:15.4 & 2-3 May 2016 & J  & 4.53 & CIV   & [1], [3],[4]   \\ 
\ldots &\ldots & \ldots&\ldots &    31 Jul 2017 & J & 1.87 & CIV & [1], [3], [4]  \\
EGS-zs8-2 &   7.477   & 14:20:12.09  &  +53:00:27.0 & 2-3 May 2016 & J & 4.53 &  CIV & [1], [4]  \\ 
\ldots & \ldots & \ldots&\ldots &    31 Jul 2017  & J & 1.87 & CIV & [1], [4] \\
  A1703-zd6 &  7.045   &    13:15:01.01 & +51:50:04.3 & 29 Apr 2016  & H & 2.47 & CIII]  & [5],[6] \\
  A383-2211 & 6.031 & 02:48:01.39 & -03:32:58.4 & 15,18,25 Dec 2013 & VIS & 4.78 & Ly$\alpha$ & [7] \\
 \ldots &  \ldots  &  \ldots  & \ldots &  \ldots  & NIR &  \ldots  & CIV, He II, OIII] , CIII] & [7] \\

  Abell2218-S3.a & 5.576   &  16:35:51.96 &  +66:12:45.9  & 08 Sep 2015  & J & 1.33 & CIII]   & [8] \\
          \ldots & \ldots & \ldots&\ldots   & 29 Apr 2016  & J & 1.20  & CIII]   & [8] \\
   Abell2218-S3.b    & 5.576   &  16:35:52.08 &  +66:12:51.8  & 08 Sep 2015  & J & 1.33 & CIII]   & [8] \\
          \ldots & \ldots & \ldots&\ldots  & 29 Apr 2016   & J &1.20 & CIII]   & [8] \\
   J14144+5446   &  5.426   &  14:14:46.82 & +54:46:31.9  & 07 Apr 2015   & J & 1.08 & CIII]  & [9] \\ \hline \hline
  &&&& Photometric Targets&&&& \\  \hline  \hline
  A1703-zD4 & 8.4$^{+0.9}_{-1.4}$ &13:15:07.19 & 51:50:23.5 & 29 Apr 2016  & H & 2.47 & CIII]  & [10] \\
  A1703-zD1 & 6.7$^{+0.2}_{-0.1}$ & 13:14:59.42 & 51:50:00.8 & 29 Apr 2016  & H & 2.47 & CIII]  & [10] \\
  Abell2218-C1.b & 6.7$^{+0.1}_{-0.1}$ & 16:35:54.40 & 66:12:32.8 & 29 Apr 2016   & J &1.20 & CIV, He II, OIII]   & [11] \\
  Abell2218-C1.c & 6.7$^{+0.1}_{-0.1}$ & 16:35:48.92 & 66:12:02.4 & 29 Apr 2016   & J &1.20 & CIV, He II, OIII]   & [11] \\  \hline \hline
   \end{tabular}
   
\caption{Galaxies targeted in this study. A383-2211 was observed with the VLT/X-shooter, J14144+5446 with the LBT/LUCI, and the rest were observed with the Keck/MOSFIRE. The final column provides the reference to the article where the objects were originally identified. References: [1] \citet{RobertsBorsani2016}; [2] \citet{Zitrin2015}; [3] \citet{Oesch2015};
 [4] \citet{Stark2017}; [5] \citet{Schenker2012}; [6] \citet{Stark2015b}; [7] \citet{Bradley2014}; [8] \citet{Ellis2001} ; [9] \citet{McGreer2017} ; [10] \citet{Bradley2012}; [11] \citet{Kneib2004}}
\end{table*}

We have obtained near-infrared spectroscopic observations of four spectroscopically confirmed Ly$\alpha$ 
emitting galaxies at $z>7$ and an additional three at $5.4<z<6.0$.  We also have obtained deep spectra of 
three bright $z\simeq 7$ sources  lacking Ly$\alpha$ emission.  
Observations have been obtained with Keck, VLT, and the LBT.  
Below we describe the observational setup and data reduction procedure.  

\subsection{Keck/MOSFIRE spectroscopy}

The majority of spectra presented in this paper were obtained using the multi-object spectrograph 
MOSFIRE \citep{McLean2012} on the Keck I telescope.  Over five different observing runs between 
2015 and 2017, we targeted $z\simeq 6-9$ galaxies in the EGS, Abell 1703, and Abell 2218 fields.
Details of the observations are summarized in Table 1.

The spectra constrain the strength of He II and  UV metal emission lines in known $z>7$ 
galaxies.  Our target selection is driven by the need for sources to have precise redshifts, 
bright continuum flux, and redshifts which 
place  at least one of the strong UV lines at a wavelength with significant atmospheric transmission.   
We designed five separate masks which are described in more detail below.  For each mask, 
we included 1 or 2 isolated stars to monitor seeing and to calculate the absolute flux calibration.  
Each mask is constructed with slit widths of 0\farcs7.   Below we briefly describe the observing conditions 
and key targets.  More details on the physical properties of the sources will be provided in \S3.

The EGS CANDELS area contains three of the thirteen  $z>7$ galaxies with S/N$>$5 Ly$\alpha$ 
detections.   We designed two masks to cover the three targets.  
The first mask was focused on EGSY8p7, the $z=8.68$ 
galaxy confirmed in \citet{Zitrin2015}.   At this redshift, we are able to probe CIV, He II, and 
OIII] in the H-band.  The CIII] line is situated between the H and K-bands in a region of poor 
atmospheric transmission.  We obtained 5.93 hrs of on-source integration in H-band on the mask 
on 29-30 April 2016.  Conditions were clear with seeing of 0\farcs8.   The mask  was filled with 
lower redshift ($z\simeq 1-2$) sources picked to have extreme equivalent width rest-frame optical 
emission lines in the H-band.  We will discuss the spectra of the filler sources in a separate paper  
(Tang et al. 2018, in prep).

The other two $z>7$ Ly$\alpha$ emitters in the EGS field, EGS-zs8-1 ($z=7.73$) and EGS-zs8-2 ($z=7.47$), 
are close enough to fit on the same MOSFIRE mask.  We previously obtained MOSFIRE H-band observations 
of both systems, revealing strong CIII] emission in EGS-zs8-1 and no CIII] emission in EGS-zs8-2 \citep{Stark2017}.  
Here we present new J-band observations aimed at constraining the strength of CIV emission in both galaxies.  
The OIII] and He II emission lines are situated between the J and 
H-bands where atmospheric transmission is low.   As with our other EGS mask, we included $z\simeq 1-2$ 
extreme emission line galaxies as filler targets.   We obtained a total of 6.4 hours of integration in J-band, with 4.53 
hours secured over 2-3 May 2016 and an additional 1.87 hours on 31 July 2017.   Conditions 
were clear for both runs with seeing of 0\farcs7 (May 2016) and 0\farcs9 (July 2017).

The Abell 1703 field has several bright gravitationally lensed $z\gtrsim 7$ galaxies that are ideal for 
spectroscopic study \citep{Bradley2012}, including a Ly$\alpha$ emitter at  z$_{\rm{Ly\alpha}}$=7.045 
(A1703-zd6; \citealt{Schenker2012}) and six additional targets with photometric redshifts between $z\simeq 6.4$ and 8.8.
We previously targeted the field in J-band, revealing nebular CIV emission in A1703-zd6 \citep{Stark2015b}. 
We now target CIII] emission in A1703-zd6 with H-band observations of the Abell 1703 field.   We 
also include the photometric $z\gtrsim 7$ targets A1703-zd1 and A1703-zd4 on the mask, 
with the goal of constraining CIII] in the former and CIV, He II, OIII], and CIII] in the latter.  
Empty real estate on the mask was filled with targets detected with Herschel  (Walth et al, in prep).  
Conditions were clear with seeing of 0\farcs7 throughout the 2.47 hours of on-source integration time 
on 29 April 2016.   

The Abell 2218 cluster field has two relevant sources for UV metal line follow-up: A2218-S3, a multiply-imaged 
Ly$\alpha$ emitter at  z$_{\rm{Ly\alpha}}$=5.576 \citep{Ellis2001} and A2218-C1, a bright triply-imaged $z\simeq 6.7$ galaxy lacking
Ly$\alpha$ emission \citep{Kneib2004,Egami2005}.  We designed a mask that includes the two images of the  z$_{\rm{Ly\alpha}}$=5.576
galaxy (S3a, S3b) and two of the images of the $z\simeq 6.7$ source (C1b, C1c).   We filled the mask with other lensed  
galaxies in the field: C2.a ($z\simeq 3.104$; \citealt{Richard2011a}), C3.a ($z\simeq 2.8$; \citealt{Eliasdottir2007}), S2.1.a ($z=2.51$; \citealt{Ebbels1996}), and S6.a ($z=1.03$; \citealt{Kneib1996}).  
We observed the mask for a total of 1.2 hours in the J-band on 29 April 2016. For the z$_{\rm{Ly\alpha}}$=5.576 galaxy,
we also added data from a previous program (PI: Zitrin) who observed the galaxy for 1.33 hours on 08 September 2015.
Both nights were clear with seeing of 0\farcs9  (September 2015) and 0\farcs8 (April 2016).

Data were reduced using the publicly-available MOSFIRE Data Reduction Pipeline. The pipeline performs flat fielding, 
wavelength calibration and background subtraction. The reduction process outputs two dimensional reduced spectra along 
with a signal-to-noise map and error map for each slit on the mask.   Finally, one-dimensional spectra were obtained using a 
boxcar extraction with an aperture size of 1.5$\times$ observed seeing, typically 6-8 pixels (1\farcs08 - 1\farcs44).
The flux calibration was performed in two stages. The telluric correction was first applied using longslit observations of a 
spectrophotometric standard star that we targeted prior to the science observations.  The absolute flux calibration was 
then calculated using the known fluxes and observed counts of the isolated stars on the mask.
The spectral resolution is calculated by fitting the width of isolated sky lines.  For our chosen slit widths, 
we derive an average FWHM of 3.8 \AA\ in the J-band and 4.7 \AA\ in the H-band.

\subsection{VLT/X-Shooter spectroscopy}

One additional source (A383-2211) was observed as a part of programme 092.A-0630(A) (PI: Richard) with the instrument X-Shooter \citep{Vernet2011} on the 
Very Large Telescope (VLT). 
The observations were performed on the nights of 15, 18, and 25 December 2013 for 6 observing blocks (OBs) of 1 hour each. Each OB comprised 
three exposure of 955 seconds in the visible (VIS) arm and three exposures of 968 seconds in the near infrared (NIR) arm. The VIS arm provided wavelength coverage from
 5336~\AA\ to 10200~\AA\  at a spectral resolution of 1.1~\AA, whereas the NIR arm provides coverage  from 9940~\AA\
to 24750~\AA\ at a spectral resolution of 2.1~\AA. We used a slit of $11\farcs0\times0\farcs9$ size 
oriented at a position angle of 90\textdegree. 
The observations were conducted using a dither sequence of $\pm$2\farcs5 along the slit.
The conditions were clear with an average seeing of 0\farcs6.

The data reduction was performed using version 2.2.0 of the X-Shooter pipeline. The details of the 
procedure were presented in \citet{Stark2015a}.   The pipeline outputs 18 reduced exposures which we then  combine using standard IDL and IRAC routines. 
A spectroscopic standard star was observed on the same 
night to calibrate flux of the spectrum. Multiple telluric stars were observed to 
calculate a median telluric correction for telluric absorption in the NIR arm.

\subsection{LBT/LUCI spectrosocopy}

We observed  the $z_{\rm{Ly\alpha}}=5.426$ Ly$\alpha$ emitter J141446.82+544631.9 (hereafter J1414+5446; \citealt{McGreer2017}) 
on 2015 April 07 for 1.08 hours using the LUCI near-IR spectrograph on the Large Binocular Telescope (LBT).  
We used the N1.8 camera and 210\_zJHK grating in longslit mode.   The observations were conducted in the J-band with the goal of constraining the strength of 
the CIII] emission line.  A slit width of 1\farcs0 was used, resulting in a spectral resolution of 2.93 \AA. 
A bright reference star (J=17.9) was placed along the slit to monitor seeing and calibrate the absolute flux scale.  
Data were taken with a standard ABBA dither sequence with a 7\farcs0 offset between the A and B positions.  Conditions 
were clear with seeing of 1\farcs0 throughout the observations.

The reduction was performed using an IDL long slit reduction package for LUCI (see \citealt{Bian2010} for details).  
The package provides flat fielding and sky subtraction using standard routines.  The wavelength solution was  
calculated using night sky lines. The software outputs a fully reduced 2D spectrum which we visually examined 
for faint emission lines.   We identified the spatial position of the galaxy on the spectrum using the 
trace of the reference star and the known angular offset between the star and galaxy.   
A one dimensional spectrum was created using a boxcar extraction with a 6 pixel (1\farcs5) aperture.   
The extracted spectrum was corrected for telluric absorption using the A0V star spectrum, and the absolute 
flux scale of the extracted spectrum was then derived using the flux of the reference star on the slit.   

\section{Results} \label{sec:result}

We describe new UV spectral line constraints for the targets presented in \S2. 
For each source, we first summarize known 
physical properties and then detail the results of the newly-obtained spectra.  

\subsection{EGSY8p7}
EGSY8p7 is a bright (H$_{160}$=25.3) galaxy identified by RB16 and confirmed 
spectroscopically by \citet{Zitrin2015} via detection of Ly$\alpha$ emission at 
$z_{\rm{Ly\alpha}}=8.683$ in a J-band spectrum obtained with MOSFIRE (see Figure 1).  The red IRAC 
color ([3.6]-[4.5]=0.76) is consistent with the presence of strong [OIII]+H$\beta$ 
emission in the [4.5] filter.  A combined [OIII]+H$\beta$ rest-frame equivalent width of 
$895\pm112$\AA\ is required to explain the [4.5] flux excess relative to 
the underlying continuum \citep{Stark2017}.
 
The J-band  spectrum and the new H-band spectrum cover 11530~\AA\ to 13458~\AA\ and 
14587 ~\AA\ to 17914~\AA, respectively, corresponding to 1191~\AA\ to 1390~\AA\ and 
1506~\AA\ to 1850~\AA\ in the rest-frame.   In addition to Ly$\alpha$, this wavelength range 
allows constraints to be placed on NV, CIV, He II, and OIII].   We guide our search of 
these lines using the Ly$\alpha$ redshift and the characteristic velocity offset of Ly$\alpha$ from systemic.  
We consider Ly$\alpha$ velocity offsets between 0 and 500 km s$^{-1}$ \citep{Erb2014,Stark2017}, 
implying systemic redshifts in the range z$_{\rm{sys}}$=8.667 to 8.683 for EGSY8p7.  
Nebular He II, and OIII] are typically found to trace the systemic redshift (e.g., \citealt{Shapley2003, Stark2014, Senchyna2017}), so we predict a spectral window for each line using the range of systemic redshifts 
derived above.  CIV is a resonant transition and is found at redshifts between systemic and that of 
Ly$\alpha$ \citep{Christensen2012, Stark2014, Stark2015b, Vanzella2016,Vanzella2017, Mainali2017, 
Laporte2017,Berg2018}, so we consider a spectral window ranging from the lowest possible systemic redshift 
(z$_{\rm{sys}}$=8.667) and the Ly$\alpha$ redshift ($z_{\rm{Ly\alpha}}=8.683$).  

\begin{figure*}
\centering
\begin{tabular}{ccc}

\subfloat{ \includegraphics[scale=0.6]{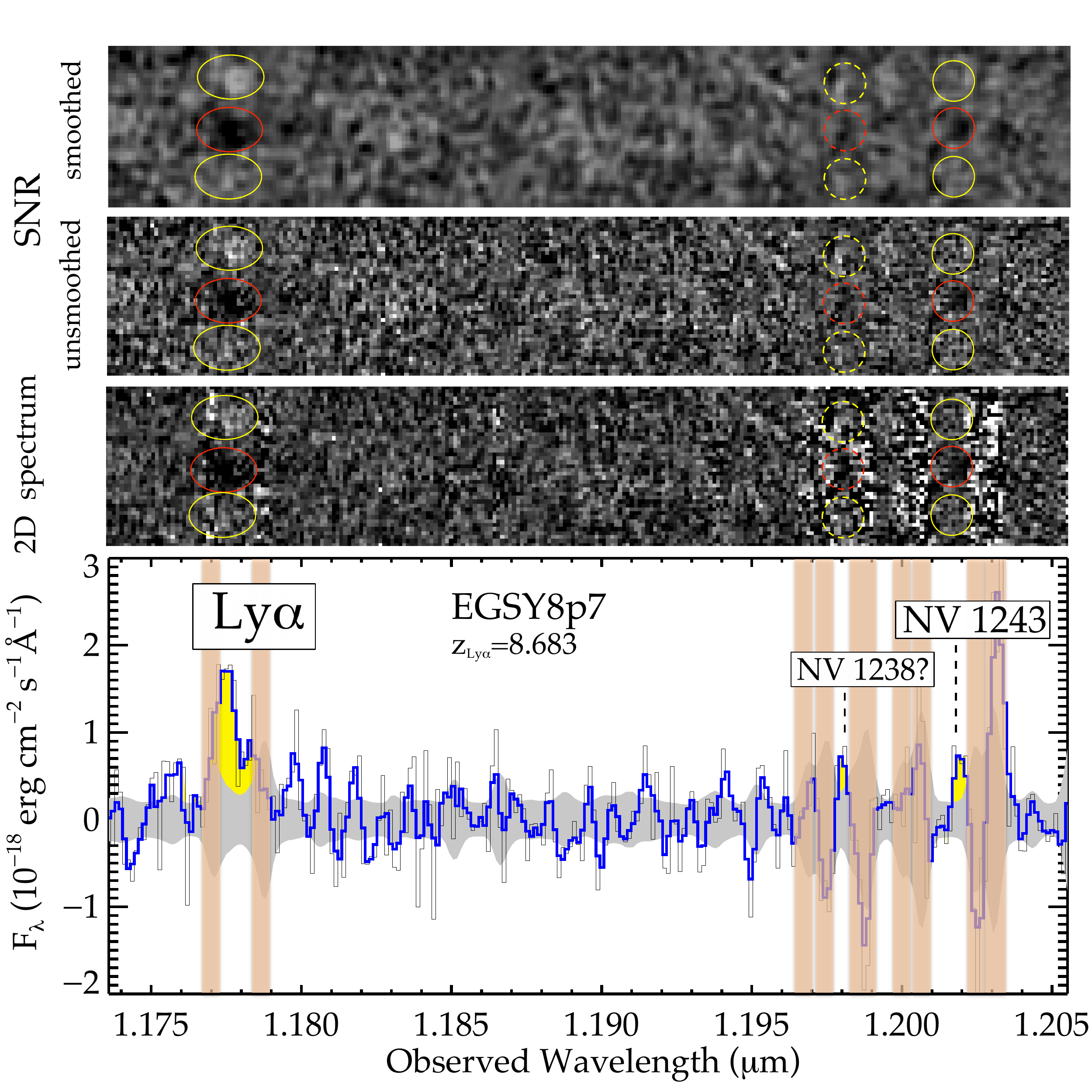}}

\end{tabular}
\caption{ Keck/MOSFIRE two-dimensional and one-dimensional J-band spectrum of EGSY8p7 similar to that  originally published in \citet{Zitrin2015}. 
(Top:) The two-dimensional SNR map (smoothed and unsmoothed) showing the NV$\lambda$1243 detection where positive fluxes are denoted by black. 
The red circle shows positive peak of the detection at the same spatial position as expected from Ly$\alpha$ emission while yellow circles 
denote negative peaks as expected per the dither sequence. The red dotted circle denote the tentative positive flux from the NV$\lambda$1238 component, while yellow dotted circles denote the location of negative peaks as per dither sequence. (Middle:) Similar figure as the top panel showing the two-dimensional spectrum. 
(Bottom:) One-dimensional spectrum showing the Ly$\alpha$ line along with the NV$\lambda$1243 and tentative NV$\lambda$1238 emission features. 
The black curve denotes the extracted 1D flux whereas the grey region indicates 1$\sigma$ noise level. 
The smoothed (2-pixel) flux is shown in blue curve.}
\label{fig:egs8p7_lyanv}
\end{figure*}

We first consider constraints on CIV, He II, and OIII] from the new H-band spectrum.  
Taking the range of possible systemic redshifts for EGSY8p7 that we derive above, we predict spectral windows for  
CIV$\lambda$1548 (14966~\AA\ to 14991~\AA), CIV$\lambda$1550 (14991~\AA\ to 15016~\AA), 
He II$\lambda$1640 (15860~\AA\ to 15885~\AA), OIII]$\lambda$1661 (16055~\AA\ to 16082~\AA), and 
OIII]$\lambda$1666 (16106~\AA\ to 16133~\AA).   As can be seen in Figure 2, each window is 
mostly ($>$80\%) free from strong skylines.  No significantly-detected emission lines are seen.  
We calculate typical 3$\sigma$ line flux limits in each spectral window by summing the error spectrum in quadrature over 10~\AA\ ($\sim$200 km s$^{-1}$).   
As this is at the upper bound of line widths found for UV nebular emission lines other than 
Ly$\alpha$ at these redshifts (e.g., \citealt{Stark2014, Bayliss2014, James2014, Mainali2017}), 
it provides a conservative estimate of our sensitivity to emission lines.   To calculate the 
limits on the equivalent width, we compute the continuum flux 
density expected at the wavelength of each spectral line using the broadband SED fit in our 
previous studies (e.g., \citealt{Zitrin2015, Stark2017}).  The measured limits on line flux and equivalent width are presented in Table 2. The non-detections place strong 
constraints on CIV and OIII], implying that the individual components of each doublet likely have 
rest-frame equivalent widths below 4.6~\AA.  Because He II is located in a somewhat noisier region, our 
constraints are less stringent.  The non-detection implies a rest-frame equivalent width below 14.8~\AA\, provided the He II feature is not obscured by the single skyline in the predicted spectral window (see Figure 2).

\begin{figure*}
\centering
\begin{tabular}{ccc}
\hspace{-0.6 cm}
\subfloat{ \includegraphics[scale=0.6]{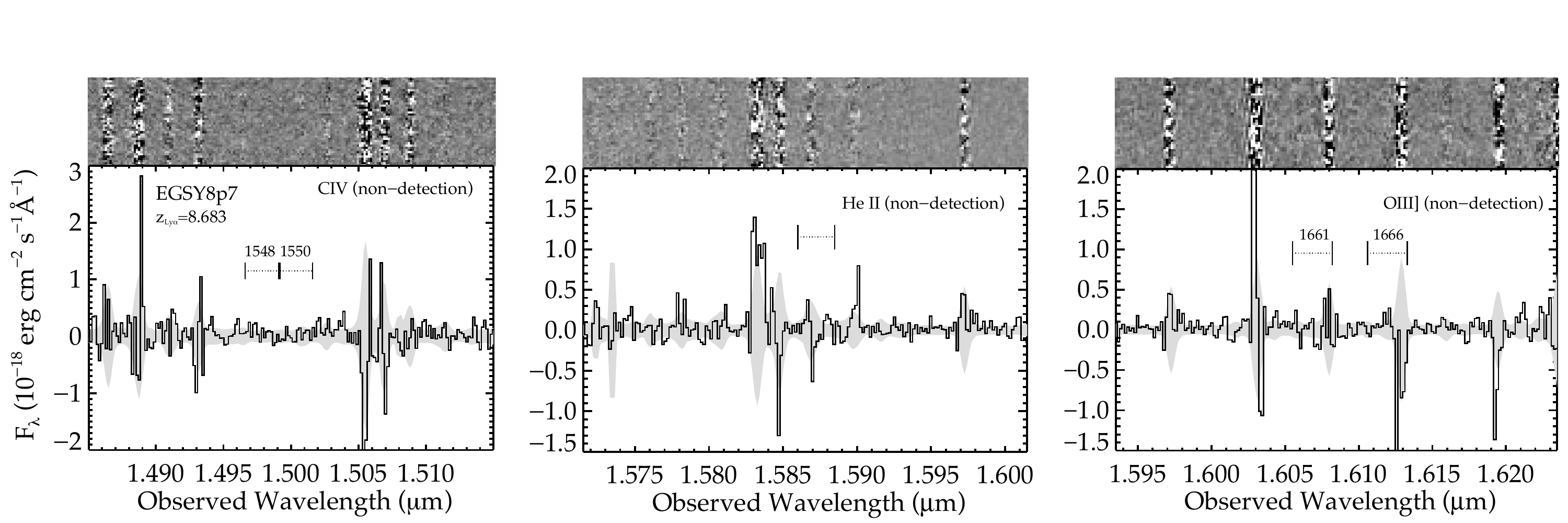}}

\end{tabular}
\caption{ MOSFIRE 2D spectrum (top) and 1D spectrum (bottom) of EGSY8p7 showing the spectral region around expected location of CIV (left), He II (middle) and OIII] (right). 
The black curve denotes the extracted 1D flux whereas the grey region indicates 1$\sigma$ noise level. The dotted black lines represent the spectral window where we expect the relevant lines to fall. See \S3.1 for details. }
\label{fig:egs8p7}
\end{figure*}

We now consider constraints on nebular NV in the J-band discovery spectrum from \citet{Zitrin2015}.  
For redshifts between  z$_{\rm{sys}}$=8.667 and  z$_{\rm{sys}}$=8.683, we 
expect NV$\lambda$1238 and NV$\lambda$1243 to lie between 11975 and 11995 ~\AA\ and 
12017 to 12037~\AA, respectively.   An emission feature (S/N=4.6) is seen centered at 12019.5~\AA\ (Figure 1), 
within the window where NV$\lambda$1243 would be expected.   The feature is at the exact spatial position of 
the Ly$\alpha$ line and is adjacent to but cleanly separated from skylines. The line is unresolved in the MOSFIRE spectrum (FWHM$<$90 km s$^{-1}$). We find that an aperture 
of 6 pixels ($\sim2\times$ the spectral resolution) maximizes the S/N. We directly integrate the object and error spectrum (the latter in the quadrature) over this window, revealing a line flux of  $2.8\times$10$^{-18}$ 
erg s$^{-1}$ cm$^{-2}$ and an error (1$\sigma$) of  $0.6\times$10$^{-18}$ erg s$^{-1}$ cm$^{-2}$.
If the line is NV$\lambda$1243, the blue component of the doublet would be situated near two strong skylines (Figure 1) 
making detection difficult.   Nevertheless we do detect positive emission at the precise wavelength predicted for the 
NV$\lambda$1238 feature (11981~\AA), but the skylines prohibit accurate flux measurement.

\begin{figure}
\centering
\begin{tabular}{c}
\hspace{-0.3 cm}
\vspace{-0.5 cm}
\subfloat{ \includegraphics[angle=90,scale=0.55]{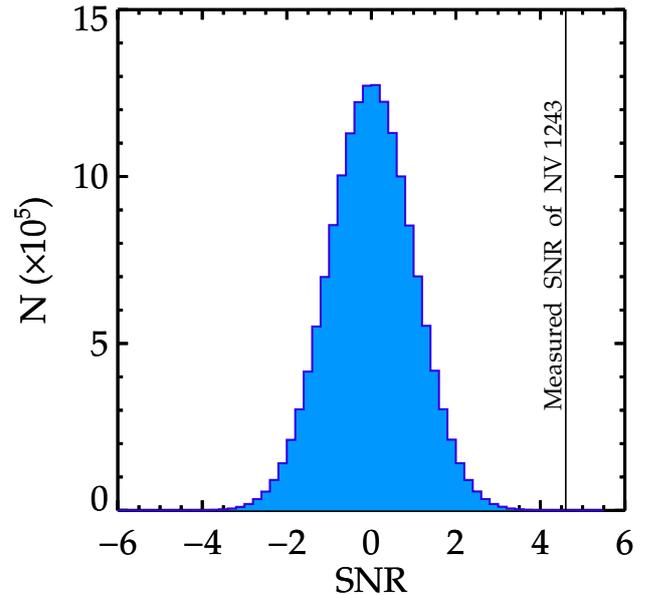}}

\end{tabular}
\caption{The signal-to-noise (SNR) distribution in 10$^6$ realizations of the error spectrum within the 16 pixel window defined by the Ly$\alpha$ (see \S3.1). 
Apertures of  size 6$\times$6 pixel, similar to the one used to measure the NV detection, were used to estimate the SNR. 
The vertical black line denotes the detected significance of the NV 1243 line.}

\end{figure}

The odds of finding a 4.6$\sigma$ feature at the spatial position of Ly$\alpha$ and 
within the narrow 500 km s$^{-1}$ (16 pixel) spectral window defined above are exceedingly low.   We 
can test this by generating a large number (10$^6$) of realizations of the error spectrum in the 16 pixel 
box surrounding the putative NV$\lambda$1243 detection.  We find features with S/N in excess of 4.6 in 
only 32 of the 10$^6$ realizations, implying a very low probability (0.003\%) that the line is not real (Figure 3).  
We further investigate the likelihood of the detection by looking at the significance of the emission in 
subsets of the stacked spectrum.  The J-band data was obtained over two nights (see \citealt{Zitrin2015} for details) 
with integration times of 158 min (night one) and 128 min (night two) contributing to the total stack. 
The  data from the first night were obtained under significantly better conditions (seeing of 0\farcs6) than  
 the second night (0\farcs8).  As reported in \citet{Zitrin2015}, the Ly$\alpha$ detection significance is higher 
in the night one stack (S/N=6.0) than in the night two data (S/N=3.8), consistent with expectations given  
the longer integration time and improved seeing.   
The significance of the NV$\lambda$1243 feature is also found to be larger on night 1 (S/N=4.1) relative 
to night 2 (S/N=2.0).   The increase in significance between night two and night one is consistent with 
that measured for Ly$\alpha$.  The fact that the significance of the feature scales with atmospheric 
conditions in a manner consistent with Ly$\alpha$ further bolsters confidence that the NV detection 
is real and associated with EGSY8p7.   

The rest-frame equivalent width implied by the NV$\lambda$1243 detection is 4.2$\pm$0.9~\AA. 
Here we derive the continuum flux (and associated error) from the broadband SED, as is standard at these redshifts.
We include the error in the underlying continuum and line flux in the equivalent width, and we take care to 
not include filters contaminated by nebular emission in our derivation of the continuum flux. The uncertainty in the equivalent width is dominated by the line flux uncertainty which is $22\%$. Using 
the standard theoretical flux ratio for the NV doublet (NV$\lambda$1238/NV$\lambda$1243=2; \citealt{Bickel1969,TorresPeimbert1984}), we 
can estimate the total flux (8.4$\pm1.8\times$10$^{-18}$ erg cm$^{-2}$ s$^{-1}$) 
and rest-frame equivalent width (12.6$\pm$1.8~\AA).   The observed wavelength of the likely 
NV$\lambda$1243 feature would indicate a redshift of $z=8.671$, implying a 
Ly$\alpha$ velocity offset (from NV) of 362 km s$^{-1}$.  While NV is a resonant transition, 
it has been shown to trace the systemic redshift to within 20 km s$^{-1}$ in another of the 
$z>7$ Ly$\alpha$ emitters \citep{Laporte2017, Pentericci2016}.  If the likely NV feature 
also traces the systemic redshift in EGSY8p7, it would indicate a large Ly$\alpha$ velocity offset, 
consistent with other similarly massive sources at high redshift (e.g. \citealt{Erb2014, Stark2017}).   
Such a large velocity offset may aid in the escape of Ly$\alpha$ through the IGM \citep{Mason2017}.

\subsection{EGS-zs8-1 and EGS-zs8-2}

Our second mask in the EGS contained  EGS-zs8-1 and EGS-zs8-2 from RB16.   Both are bright 
(H$_{\rm{160}}$=25.0 and 25.1, respectively) and have red IRAC [3.6]-[4.5] colors that 
suggest strong [OIII]+H$\beta$ emission (EW=911$\pm$122~\AA\ and 1610$\pm$302~\AA\ rest-frame, 
respectively \citep{Stark2017}.   Detections of Ly$\alpha$ in EGS-zs8-1 and EGS-zs8-2 confirm 
their redshifts to be $z_{\rm{Ly\alpha}}$=7.730 \citep{Oesch2015} and 7.477 (RB16, \citealt{Stark2017}).   
H-band MOSFIRE observations presented in \citet{Stark2017} 
revealed CIII] emission in EGS-zs8-1  (rest-frame EW of $22\pm2$~\AA) and a non-detection in 
EGS-zs8-2.  The latter provides a 3$\sigma$ upper limit on the total CIII] EW in EGS-zs8-2 of $14.2$~\AA.  
Our new J-band spectra allow us to constrain the strength of CIV emission in both sources.   As 
with EGSY8p7, we predict the wavelength of CIV using estimates of the systemic redshift.
The systemic redshift of EGS-zs8-1 ($z_{sys}$=7.723) is determined using the wavelength of the [CIII], 
CIII] doublet.  For EGS-zs8-2 we consider systemic redshifts in the range $z_{sys}$=7.463 to 7.477, following 
the same approach as we took for EGSY8p7.   

We consider NV redshifts between that of systemic and Ly$\alpha$.  For EGS-zs8-1, this implies 
a spectral window of 10797~\AA\ to 10815~\AA\  (NV$\lambda$1238) and 10797~\AA\ to 10815~\AA\ 
(NV$\lambda$1243).  These wavelengths are covered in the discovery Y-band MOSFIRE spectrum 
presented in \citet{Oesch2015}.  
The  NV$\lambda$1238 window is completely clear of skylines, while 37\% of the NV$\lambda$1243 
window is obscured.  No emission lines are detected in either window, implying 3$\sigma$ upper limits 
on the equivalent width of NV$\lambda$1238 ($<$7.4~\AA) and NV$\lambda$1243 ($<$29.6~\AA).  
As  NV$\lambda$1238 is the stronger of the two components, its absence in the Y-band spectrum 
confirms that NV is not present in EGS-zs8-1 at the same intensity as it is found in EGSY8p7.
The redshift of EGS-zs8-1 makes detection of CIV  difficult.  The J-band spectrum covers 
11530~\AA\ to 13520~\AA, corresponding to rest-frame wavelengths between 1320~\AA\ and 
1549~\AA. This places CIV$\lambda$1548 at the red edge of the J-band where sensitivity is 
reduced and cuts out the $\lambda$1550 component entirely.  To predict the observed 
wavelength of CIV$\lambda$1548, we again consider CIV redshifts between that of 
systemic and Ly$\alpha$.  This translates into a spectral window spanning 13493~\AA\ 
to 13515~\AA. No emission line is visible. For a  10~\AA\ line width, we derive a 3$\sigma$ 
limiting flux (rest-frame EW) of 1.1$\times$10$^{-17}$ erg cm$^{-2}$ s$^{-1}$ (17~\AA) 
for CIV$\lambda$1548.    We can use the theoretical line ratio of CIV doublet components 
(CIV$\lambda$1548/CIV$\lambda$1550=2) to derive an upper limit on the total CIV doublet 
strength (1.7$\times$10$^{-17}$ erg cm$^{-2}$ s$^{-1}$), implying a rest-frame EW below 29~\AA.  
The CIV non-detection places a 3$\sigma$ upper bound on the CIV/CIII] flux ratio in EGS-zs8-1 of less than 2.1.  
This is  consistent with expectations for nebular lines powered by massive stars and AGN 
(e.g., \citealt{Feltre2016, Jaskot2016,Byler2018}).  The lack of atmospheric absorption will enable 
{\it JWST} to  extend these UV line measurements to deeper limits.  

The J-band spectrum of EGS-zs8-2 (Figured 4a) probes rest-frame wavelengths between 1360~\AA\ to 1596~\AA, 
allowing constraints to be placed on the strength of nebular CIV.  The absence of strong CIII] in EGS-zs8-2 
could potentially reflect a hard ionizing spectrum that produces a large CIV/CIII] ratio.  
Following the same approach as above, we consider a window between 13102~\AA\ to 13124~\AA\ 
for CIV$\lambda$1548 and 13124~\AA\ to 13146~\AA\  for CIV$\lambda$1550.  The CIV$\lambda$1548 
window is free of strong skylines, but no convincing nebular emission is apparent.   
The non-detection of CIV$\lambda$1548 in this window allows us to place a 
3$\sigma$ upper limit of  1.2$\times$10$^{-18}$ erg cm$^{-2}$ s$^{-1}$ and 2.4~\AA\ for the line flux 
and rest-frame equivalent width, respectively.   There is a skyline that covers 21\% of the CIV$\lambda$1550 
spectral window.  While the red component of the doublet could 
be obscured by this sky feature, the blue component (stronger by a factor of two) would have been 
visible if CIV emission was strong.  The non-detection of the doublet suggests that 
the total equivalent width of nebular CIV is below 4.8~\AA\ (3$\sigma$) for EGS-zs8-2. 
We also consider the presence of NV in the Y-band Keck/MOSFIRE spectrum presented in \citet{Stark2017}. 
No emission lines are seen in the predicted windows for NV$\lambda$1238 (10484~\AA\ to 10501~\AA) 
and NV$\lambda$1243 (10518~\AA\ to 10535~\AA).   As both of these wavelength ranges are mostly 
free of strong skylines, the non-detection provides a 3$\sigma$ upper limit to the rest-frame equivalent width of 
NV$\lambda$1238 ($<$2.2~\AA) and NV$\lambda$1243 ($<$4.5~\AA).  EGS-zs8-2 is 
thus the only of the four RB16 sources lacking a UV metal line detection, likely implying a less extreme 
radiation field. 

\subsection{Abell 1703}

The top priority target on our Abell 1703 mask was A1703-zd6, a bright gravitationally lensed galaxy (H$_{160}$=25.9) 
first identified by \citet{Bradley2012}. Lyman-$\alpha$ emission was detected at 9780~\AA, indicating a redshift 
of  $z\rm_{Ly\alpha}$=7.045 \citep{Schenker2012}.  After correcting for cluster magnification ($\mu$=5.2), the 
absolute magnitude of A1703-zd6 is relatively faint (M$_{\rm{UV}}$=$-$19.3), providing a glimpse of the properties 
of a lower mass reionization-era system with visible Ly$\alpha$ emission.  A J-band spectrum obtained with MOSFIRE revealed 
strong CIV$\lambda\lambda$1548,1550 emission (EW=38~\AA\ rest-frame) \citep{Stark2015b}, requiring 
a hard ionizing spectrum with 48 eV photons capable of triply ionizing carbon.  Since CIV is a resonant line, we cannot use it to determine 
the systemic redshift. Using the same approach as for the EGS targets, we predict that the 
systemic redshift should be in a window spanning from $z_{sys}$=7.032 to 7.045.  

The new H-band spectrum of A1703-zd6 (Figure 4b)  probes 14560~\AA\ to 17880~\AA, corresponding to 1813~\AA\ and 2214~\AA\ in the 
rest-frame, allowing constraints to be placed on the strength of CIII] emission. By constraining the line ratio 
of CIV/CIII], we hope to better constrain the shape of the ionizing spectrum powering the intense nebular emission.
The range of possible systemic redshifts situates  [CIII]$\lambda$1907 
at 15313~\AA\ to 15339~\AA\ and CIII]$\lambda$1909 at 15330~\AA\ to 15356~\AA.  
We detect no emission line in either window.  While there is a skyline that covers 18\% of each window, there is 
no configuration where both the $\lambda$1907 and $\lambda$1909 components are 
obscured.  If one line is obscured, then the other component should be in a clean part of the spectrum.  
We calculate a 3$\sigma$ limiting flux (equivalent width) of 3.3$\times$10$^{-18}$ erg cm$^{-2}$ s$^{-1}$ (19.8~\AA) 
for each component of the CIII]$\lambda\lambda$1907,1909 doublet, where we have again assumed 
a 10~\AA\ FWHM for the line.  The absence of CIII] suggests that A1703-zd6 has a CIV/CIII] line ratio of 
$>$1.2 (3$\sigma$), consistent with the CIV/CIII] flux ratios (1.3-6.6) seen in AGN at lower redshifts 
(e.g., \citealt{Dors2014,Feltre2016}).  Metal poor galaxies with nebular CIV emission have also 
been observed with CIV/CIII] ratios in excess of 1.0 (e.g., \citealt{Vanzella2016,Vanzella2017, Mainali2017}), 
so the current limits do not rule out massive stars as a powering mechanism.   Deep constraints on He II and 
NV provide the best path toward determining the origin of the CIV emission. 

The MOSFIRE mask in Abell 1703 also allows us to place constraints on UV emission lines in 
two $z\gtrsim 6$ targets that lack spectroscopic redshifts.  A1703-zd1 is one of the 
brightest (H=24.0) galaxies at $z\simeq 7$.  Using broadband photometry from 
{\it HST}, \citet{Bradley2012} derived a photometric redshift of $z=6.7^{+0.2}_{-0.1}$.  The 
galaxy is highly-magnified ($\mu=9.0$) implying an absolute UV magnitude of M$_{\rm{UV}}=-20.5$. 
Deep z and J-band spectra have failed to reveal any emission lines in A1703-zd1, providing 
upper limits on the flux of Ly$\alpha$ \citep{Schenker2012} and CIV, He II, and OIII] \citep{Stark2015b}.  
By obtaining an H-band spectrum which covers 14500~\AA\ to 17800~\AA, we now constrain the 
strength of CIII] emission.  
The range of possible photometric redshifts places  [CIII]$\lambda$1907 between 
 14493~\AA\ and 15065~\AA\ and CIII]$\lambda$1909 between 14508 and 15081~\AA. 
The region is largely free of OH lines, with 59\% of the wavelength 
range having a 5$\sigma$ flux limit of 5.3$\times$10$^{-18}$ erg cm$^{-2}$ s$^{-1}$. This corresponds to a rest-frame equivalent width of 6.8~\AA\ 
for individual components of the [CIII], CIII] doublet after an aperture correction of 1.09$\times$ is applied to the line flux to account for the slit losses from this extended arc.

We apply a similar analysis to A1703-zd4, another bright (H$_{\rm{160}}$=25.4) galaxy with a 
photometric redshift of $z=8.4$.  The photometry allows a broader range of redshifts for this target, 
with acceptable solutions between $z=7.0$ and $z=9.3$ \citep{Bradley2012}.   
The source is magnified by $\mu=3.1$ implying an absolute UV magnitude of M$_{\rm{UV}}$=$-$20.6.  
The J-band spectrum from \citet{Stark2015b} constrains Ly$\alpha$ over $8.5<z<10.1$ and CIV over 
$6.4<z<7.7$.   The absence of emission lines in the J-band suggests a 5$\sigma$ EW limit of 6~\AA\ for 
 Ly$\alpha$ and CIV lines situated between OH emission lines.  The newly-acquired H-band spectrum 
samples wavelengths between 14810~\AA\ and 18140~\AA, and is sensitive to CIV at redshifts 
$8.57<z<10.69$, He II at $8.03<z<10.05$, OIII] at $7.89<z<9.92$, and CIII] at $6.76<z<8.34$.   No 
emission is visible throughout the spectrum.   For regions between sky lines (60\% of the spectrum), we 
measure a 5$\sigma$ flux limit of 5.7$\times$10$^{-18}$ erg cm$^{-2}$ s$^{-1}$.  This implies a  
rest-frame equivalent width limit of 20~\AA\ for lines located in clean regions of the J-band spectrum.

\begin{figure*}
\centering
\begin{tabular}{ccc}

\hspace{-0.3 cm}
\vspace{-1.2 cm}
\subfloat{ \includegraphics[scale=0.72]{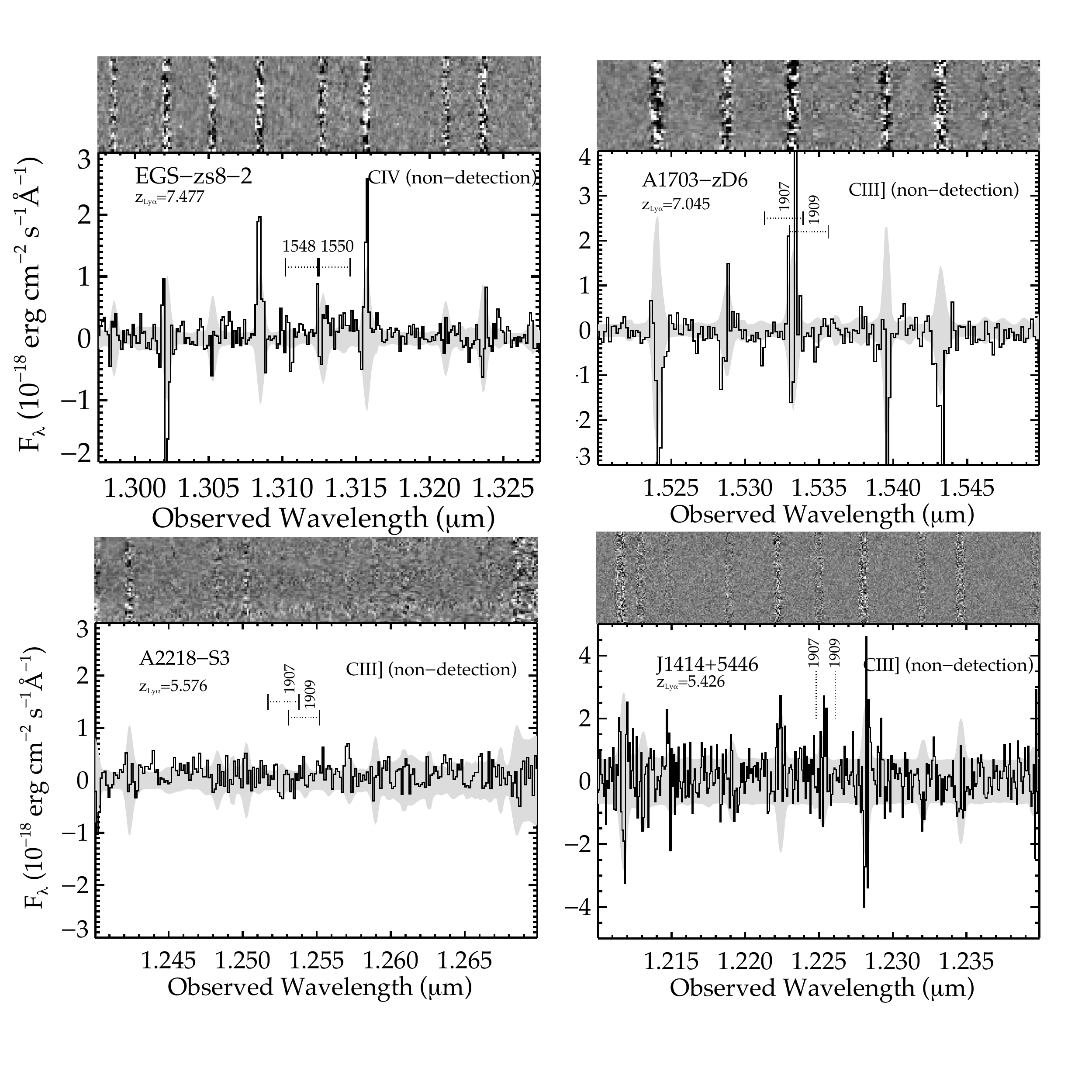}}

\end{tabular}
\caption{ 2D spectrum and 1D spectrum of EGS-zs8-2 (upper left), A1703$\_$zD6 (upper right), Abell2218-S3 (bottom left), and SDSS J1414+5446 (bottom-right) .
The black curve denotes the extracted 1D flux density whereas the grey region indicates 1$\sigma$ noise level. The dotted black lines represent the spectral window where we expect the relevant lines to fall.} 
\end{figure*}

\begin{table*}
\begin{tabular}{lccccccc}
\hline 
Object & z$_{\rm{Ly\alpha}}$ & Line &  $\lambda_{rest}$   &  Line Flux  &  Equivalent Width (W$_0$) &   Absolute Magnitude (M$\rm_{UV}$) \\
&&&(\AA)&(10$^{-18}$  erg  cm$^{-2}$ s$^{-1}$) &(\AA) & (AB) \\\hline \hline
&&&&Spectroscopic Targets&& \\  \hline  \hline
 EGSY8p7 & 8.683 & NV &  1242.80 & 2.8$\pm$0.6  &  4.2$\pm$0.9 & -21.9$^{+0.1}_{-0.1}$\\
 & \ldots & CIV &  1548.19 & $<$1.6&  $<$4.6 & \ldots \\
   &  \ldots &  \ldots  &  1550.77  & $<$1.6 &  $<$4.6 &  \ldots \\
  &  \ldots& He II  & 1640.52  & $<$5.3&  $<$14.8 & \ldots \\
  &  \ldots& OIII] &  1660.81  & $<$1.5&  $<$4.6 & \ldots\\
  &  \ldots& \ldots &  1666.15  & $<$1.5&  $<$4.6 & \ldots\\
EGS-zs8-1 &7.730 & NV &  1238.82 & $<$6.0&  $<$7.4 & -22.1$^{+0.1}_{-0.1}$ \\
& \ldots & NV &  1242.80 & $<$24.3&  $<$29.6 & \ldots  \\
          & \ldots & CIV &  1548.19 & $<$11.0&  $<$17 & \ldots  \\
EGS-zs8-2 &7.477 & NV &  1238.82 & $<$1.7&  $<$2.2 & -21.9$^{+0.1}_{-0.1}$ \\
& \ldots & NV &  1242.80 & $<$3.6 &  $<$4.5 & \ldots  \\
&\ldots & CIV &  1548.19 & $<$1.2&  $<$2.4 & \ldots\\
    &  \ldots& \ldots  &  1550.77  & $<$1.2 &  $<$2.4&  \ldots \\
A1703-zd6  &7.045&   [CIII] &  1906.68  & $<$3.3&  $<$19.8 & -19.3$^{+0.1}_{-0.4}$  \\
   &  \ldots& CIII] &  1908.73  & $<$3.3&  $<$19.8 &  \ldots\\
A383-2211 & 6.031 & Ly$\alpha$ & 1215.67 & 16.4$\pm$1.4 &   21.1$\pm$2.8 & -20.9$^{+0.2}_{-0.2}$ \\
& \ldots & NV &  1238.82 & $<$3.8 &  $<$4.9 & \ldots  \\
& \ldots & NV &  1242.80 & $<$3.8 &  $<$4.9 & \ldots  \\
    & \ldots & CIV &  1548.19  &  $<$2.5 & $<$4.6 & \ldots  \\
   &  \ldots& \ldots  &  1550.77  & $<$1.3 & $<$2.3 &  \ldots \\
   & \ldots & He II &  1640.52  &  $<$5.3 & $<$12.2 & \ldots  \\
   & \ldots & OIII] &  1660.81  &  $<$2.5 & $<$5.7 & \ldots  \\
   &  \ldots& \ldots  &  1666.15  & $<$4.9 & $<$11.2 &  \ldots \\
   & \ldots & [CIII] &  1906.68  &  $<$1.7 & $<$5.8 & \ldots  \\
   &  \ldots& CIII] &  1908.73  & $<$1.2 & $<$4.1 &  \ldots \\
Abell2218\_S3  & 5.576 &   [CIII] &  1906.68  & $<$3.2&  $<$15.6 & -16.9$^{+0.4}_{-0.3}$ \\
   &  \ldots& CIII] &  1908.73  & $<$3.2&  $<$15.6 &  \ldots \\
   J14144+5446 & 5.426 &   [CIII] &  1906.68  & $<$12.0 &  $<$7.3 & -21.1$^{+1.6}_{-0.1}$ \\
   &  \ldots& CIII] &  1908.73 & $<$ 8.4 &  $<$5.2 &  \ldots \\ \hline \hline
     &&&& Photometric Targets&& \\  \hline  \hline
     A1703-zd4 & 8.4$^{+0.9}_{-1.4}$ & CIV, He II, OIII],CIII] & \ldots & $<$ 5.7 & $<$20 & -20.6$^{+0.2}_{-0.5}$ \\
     A1703-zD1 & 6.7$^{+0.2}_{-0.1}$ & CIII] & \ldots  & $<$ 5.3 & $<$6.8 & -20.5$^{+0.2}_{-0.8}$ \\
      Abell2218-C1 & 6.7$^{+0.1}_{-0.1}$ & CIV, He II, OIII] &\ldots& $<$ 7.1 & $<$5.5 & -19.4$^{+0.2}_{-0.3}$ \\  \hline \hline
     
     \end{tabular}   \\ 
  
\caption{Rest-UV emission line constraints on sources presented in this paper. The equivalent-widths are given in rest-frame. The limits on line fluxes and equivalent widths are quoted as 3$\sigma$.}
\end{table*}

\subsection{Abell 2218}

A2218-S3 is a multiply imaged Ly$\alpha$ emitter at $z_{\rm{Ly\alpha}}=5.576$ \citep{Ellis2001}.  
Both images (A2218-S3a and S3b) show Ly$\alpha$ emission with an estimated rest-frame 
equivalent width of $239\pm25$~\AA.  The galaxy is fainter than the others reported in this paper  
(H$_{\rm{160}}$=26.6 and 26.7 for image a and b, respectively; \citealt{Richard2007}).   After correcting for 
the magnification of a and b ($\mu$=33.1 and 30.2), the absolute UV magnitude is found to be  
the faintest in our sample (M$\rm_{UV}=-16.9$).   The MOSFIRE J-band spectra of A2218-S3a and A2218-S3b cover  
11530 to 13520~\AA, corresponding roughly to 1753~\AA\ and 2055~\AA\ 
for both targets.  For Ly$\alpha$ velocity offsets between 0 and 500 km s$^{-1}$, we predict that 
[CIII]$\lambda$1907 would fall at 12517~\AA\ to 12538~\AA\ and CIII]$\lambda$1909 would fall at 
12531~\AA\ to 12552~\AA.   Both spectral windows are free of atmospheric features, but no
emission lines are apparent (Figure 4c).   The non-detection implies that each component 
of the [CIII], CIII] doublet has a flux less than 3.2$\times$10$^{-18}$ erg cm$^{-2}$ s$^{-1}$ (3$\sigma$).  
Given the underlying continuum flux density, this indicates the total [CIII], CIII] equivalent width is below 
15.6~\AA, ruling out the extreme metal line emission seen in many other lensed Ly$\alpha$ emitters 
(e.g., \citealt{Christensen2012, Stark2014,Vanzella2016}).  

We also obtained MOSFIRE observations of A2218-C1, a triply-imaged galaxy at $z\simeq 6.7$ that 
is among the brightest known (H$_{\rm{160}}$=23.9, 24.1, 25.8 for images C1a, C1b, and C1c respectively) 
in the reionization era.   Our mask contained slits on C1.b and C1.c.   The lensing configuration of the multiple 
images rules out any lower redshift interpretation of the SED \citep{Kneib2004}, providing independent verification 
that the source is at very high redshift.  Unlike most of the galaxies discussed 
in this paper, A2218-C1 does not have powerful Ly$\alpha$ emission \citep{Kneib2004, Egami2005}, 
but its brightness should allow other UV lines to be detected if they are strong.   After correcting for source 
magnification ($\mu=25$ for images a and b), the absolute UV magnitude of C1 is M$\rm_{UV}$=$-19.4$.  
The J-band MOSFIRE spectrum we have obtained covers between 11530 and 13520~\AA.  Assuming the 
source is situated at its photometric redshift ($z=6.65\pm0.1$; \citealt{Egami2005}) this wavelength range 
constrains the strength of CIV, He II, and OIII] emission.   We see no emission line in the spectrum of 
A2218-C1.b or in the fainter image A2218-C1.c.  The  5$\sigma$ line flux limit in regions between atmospheric OH 
lines ($\simeq $57\% of the spectrum) is  7.1$\times$10$^{-18}$ erg cm$^{-2}$ s$^{-1}$.  Given the 
continuum brightness of C1b, this corresponds to a rest-frame equivalent width of 5.5~\AA\ after applying an aperture correction of 1.33$\times$ to the line flux. While it 
is conceivable that one or more of the UV lines could be obscured by a sky line, the non-detection of 
any lines in the J-band (and in the z-band; \citealt{Kneib2004}) does suggest that 
A2218-C1 is not likely to be an extreme UV line emitter.   Thus far, we have yet to robustly detect intense UV metal 
lines in a $z\gtrsim 6$ source that was not also shown to have Ly$\alpha$ emission, consistent with 
the idea that the Ly$\alpha$ emitters that are currently visible at $z>7$ are those objects with the 
most extreme radiation fields.   {\it JWST} spectral observations of very bright reionization-era systems 
that lack Ly$\alpha$ should easily be able to confirm this picture. 

\subsection{SDSS J1414+5446}

The galaxy SDSS J1414+5446 is one of the brightest (i$_{\rm{AB}}=23.0$)  known at $z>5$ \citep{McGreer2017}.   
The source is thought to be lensed by a foreground cluster with a magnification factor 
in the range $5<\mu<25$  \citep{McGreer2017}, implying absolute UV magnitudes between M$_{\rm{UV}}=-21.2$ and $-19.5$.  
Extremely strong Ly$\alpha$ emission (EW=260~\AA\ rest-frame) indicates 
a redshift of $z_{\rm{Ly\alpha}}$=5.426, and the detection of NIV]$\lambda$1487  suggests a 
hard radiation field is likely present.  Based on the absence of NV and CIV, \citet{McGreer2017} conclude 
that the source is not likely to be powered by an AGN. 

Given the strong Ly$\alpha$ and NIV] emission, this object is a prime candidate for having prominent CIII] emission. 
The new J-band LUCI spectrum covers 11680 to 13195 ~\AA, providing coverage at the rest-frame wavelengths 
(1820~\AA\  to 2038~\AA) necessary to constrain the strength of CIII].   For this source, we can use the systemic redshift implied by NIV] ($z_{\rm{sys}}=5.424$) to calculate the wavelengths for [CIII]$\lambda$1907 
(12248~\AA)  and for 
CIII]$\lambda$1909 (12261~\AA).   As shown in Figure 4, the spectral windows contain a strong line which might 
affect one of the two CIII] doublet component, but not both.   No clear detection of [CIII]$\lambda$1907 or 
CIII]$\lambda$1909 is visible in the LUCI spectrum.   We estimate a 3$\sigma$ limiting flux (equivalent width) 
of 1.2$\times$10$^{-17}$ erg cm$^{-1}$ s$^{-1}$ (7.3~\AA) for [CIII]$\lambda$1907 and 8.4$\times$10$^{-18}$ 
erg cm$^{-1}$ s$^{-1}$ (5.2~\AA) for CIII]$\lambda$1909.  The absence of CIII] ($<$12.5~\AA) rules out the extreme line emission 
 seen in many other strong Ly$\alpha$ emitters at lower redshifts.   
While CIII] EW does appear to correlate with Ly$\alpha$ EW 
\citep{Shapley2003, Stark2014,Rigby2015,LeFevre2017,Maseda2017}, clearly strong metal line emission is not completely ubiquitous among the most extreme EW Ly$\alpha$ emitters.

\subsection{Abell 383-2211}

\begin{figure*}
\centering
\begin{tabular}{ccc}
\hspace{-0.9 cm}
\vspace{-0.4 cm}
\subfloat{ \includegraphics[scale=0.75]{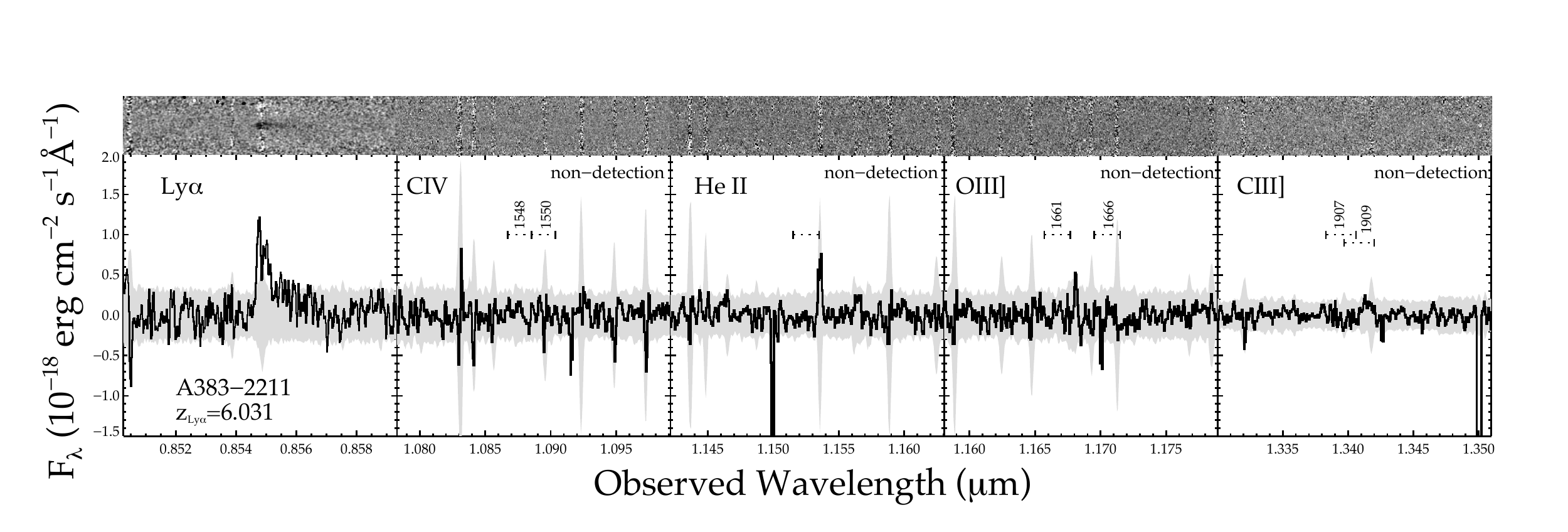}}

\end{tabular}
\caption{ X-Shooter 2D spectrum (top) and 1D spectrum (bottom) of Abell 383-2211 showing the spectral region of Ly$\alpha$, CIV, He II, OIII] and CIII], 
respectively, from left to right. Ly$\alpha$ at $z\rm_{Ly\alpha}=6.031$
is clearly visible at 0.8547$\mu$m. The black curve denotes the extracted 1D flux whereas the grey region indicates 1$\sigma$ noise level. 
The dotted black lines represent the expected spectral window where the relevant lines would fall. See \S3.5 for details.}
\label{fig:a383_lines}
\end{figure*}

A383-2211 was first identified as a dropout by \citet{Bradley2014} in deep {\it HST} imaging of the cluster Abell 383.  
The galaxy is bright (H$_{\rm{160}}$=25.2) and magnified by $\mu$=1.5.   After correcting for this magnification, the 
UV absolute magnitude is found to be M$\rm_{UV}=-20.9$, similar to that of an L$^{\star}_{\rm{UV}}$ $z\simeq 6$ galaxy 
(e.g., \citealt{Bouwens2015}).  The broadband SED suggests a photometric redshift of $z=6.0^{+0.2}_{-0.3}$ 
\citep{Bradley2014}, but prior to this work the object lacked a spectroscopic redshift.  

Our VLT/X-shooter observations of A383-2211 target Ly$\alpha$ along with NV, CIV, He II, OIII] and CIII].  
At $z\simeq 6.0$, we would expect Ly$\alpha$ in the visible arm of X-shooter.   A strong emission line is 
seen to peak at 8547.6~\AA\ (Figure 5).   The asymmetry of the observed line profile and absence of other lines in the 
VIS and NIR arms indicates this feature is Ly$\alpha$.   Adopting the wavelength of the peak flux, we
derive a redshift of $z\rm_{Ly\alpha}=6.031$, nearly identical to  the photometric redshift.   The integrated flux 
of Ly$\alpha$ is 16.4$\pm$1.4$\times$10$^{-18}$ erg cm$^{-2}$ s$^{-1}$ corresponding to a rest-frame Ly$\alpha$ 
equivalent width of 21.1$\pm$2.8~\AA.   The redshift of A383-2211 is similar to that of another lensed  Ly$\alpha$ 
emitter (A383-5.2) behind Abell 383 at $z=6.027$ \citep{Richard2011}, suggesting a likely association between the two systems.  
Strong CIII] emission has been previously detected in the $z=6.027$ Ly$\alpha$ emitter \citep{Stark2015a}; if A383-2211 
has similar stellar populations and gas conditions, we might expect to see similar metal line emission 
in the NIR arm of the X-Shooter spectrum.

We use the Ly$\alpha$ redshift  to predict the wavelengths of the fainter UV metal lines.   Allowing for 
Ly$\alpha$ velocity offsets in the range 0-500 km s$^{-1}$, we predict spectral windows for NV$\lambda$1238 (8690.0 to 8710.0), NV$\lambda$1243 (8723.5 to 8738.1), He II (11515.3 to 11534.5~\AA), 
OIII]$\lambda$1661 (11657.7 to 11677.2~\AA), OIII]$\lambda$1666 (11695.2 to 11714.7~\AA), [CIII]$\lambda$1907 
(13383.5 to 13405.9~\AA), and CIII]$\lambda$1909 (13397.9 to 13420.3~\AA).   For the resonant CIV doublet, 
we again allow redshifts between  z$_{\rm{sys}}$=6.019 and 6.031, implying a wavelength range of 10867.2 to 
10885.3~\AA\ for CIV$\lambda$1548 and 10885.3 to 10903.5~\AA\ for CIV$\lambda$1550.   
The windows specified are mostly free of atmospheric emission features (Figure 5), with OH lines taking up 
no more than 15\% of any given window.    As can be seen from Figure 5, no emission lines are present.  
The derived line flux limits are listed in Table 2.  We place 3$\sigma$ upper limits on the rest-frame 
equivalent widths of NV$\lambda\lambda$1238,1243 (4.9~\AA\ and 4.9~\AA\ ), CIV$\lambda\lambda$1548,1550 (3.8~\AA\ and 5.3~\AA), He II$\lambda$1640 (4.6~\AA), 
OIII]$\lambda\lambda$1661,1666 (3.5~\AA\ and 8.6~\AA), [CIII], CIII]$\lambda\lambda$1907,1909 (9.6~\AA\ and 12.9~\AA).  
The absence of strong CIII] suggests this object has a less extreme radiation field than the nearby galaxy A383-5.2, 
as might be expected based off of the lower Ly$\alpha$ EW.

 \begin{table*}
\begin{tabular}{lccccccc}
\hline 
Object & z$_{\rm{spec}}$ & Line &  $\lambda_{rest}$   &  Line Flux  &  W$\rm_{CIII],0}$ &   M$\rm_{UV}$ & References \\ 
&&&(\AA)&(10$^{-18}$  erg  cm$^{-2}$ s$^{-1}$) &(\AA) & (AB) &\\\hline \hline
EGS-zs8-1 &7.730 &   [CIII] &  1906.68  & 4.5$\pm$0.5&  12$\pm$2 & -22.1 & [1]\\
   &  \ldots& CIII] &  1908.73  & 3.6$\pm$0.5&  10$\pm$1 &  \ldots& [1]\\
      A1689-zD1 & 7.5 &   [CIII] &  1909  & $<$2 &  $<$4 &  -20.1 & [2]  \\
EGS-zs8-2 &7.477 &   [CIII] &  1906.68  & $<$2.3 &  $<$7.1 & -21.9 & [1]\\
   &  \ldots& CIII] &  1908.73  & $<$2.3 &  $<$7.1 &  \ldots& [1]\\
     COSY & 7.149 & [CIII] &  1906.68  & $<$0.92 & $<$3.6  & -21.8 & [3]\\
                       & \ldots   & CIII] &  1908.73  & $<$0.83 &  $<$3.2 & \ldots & [3]\\
      COSz1 & 6.854 & [CIII] &  1906.68  & $<$1.18 &  $<$3.7 & -21.6 & [3]\\
                       & \ldots   & CIII] &  1908.73  & 1.33$\pm$0.3 &  4.1$\pm$0.9 & \ldots & [3]\\
     COSz2 & 6.816 & [CIII]&  1906.68  & $<$1.57 &  $<$5.5 & -22.1 & [3]\\
                       & \ldots   & CIII] &  1908.73  & $<$1.57 &  $<$5.5 & \ldots & [3]\\    
     RXCJ2248.7-4431 & 6.110 & [CIII] &  1909  & $<$3.6 &  $<$7.9 & -20.1 & [4],[5]\\
   A383-5.2 &6.027 & [CIII]$^{a}$ &  1906.68  & 5.2$\pm$1.6 &  13.1$\pm$3.9 & -19.3 & [6]\\
                       & \ldots        & CIII] &  1908.73  & 3.7$\pm$1.1&  9.4$\pm$2.8 & \ldots & [6]\\
     A1703-23  &5.828&   [CIII] &  1906.68  & $<$1.5&  $<$1.1 & -21.7 & [7]  \\
   &  \ldots& CIII] &  1908.73 & $<$3.4 &  $<$2.6 &  \ldots & [7] \\
   Ding-3 & 5.75 &   [CIII] &  1909  & $<$3.3 &  $<$15.1 & -20.9 & [8]  \\                                                     
   Ding-1 & 5.70 &   [CIII] &  1909  & $<$5.4 &  $<$6.6 & -22.2 & [8]  \\
   Ding-2 & 5.69  &   [CIII] &  1909  & $<$3.4 &  $<$4.5 & -22.2 & [8]  \\\hline
   \end{tabular}   \\ 
   
$^{a}$Line flux  for [CIII]$\lambda$1907 is calculated assuming  [CIII]$\lambda$1907/CIII]$\lambda$1909=1.4.

\caption{Compilation of spectroscopically-confirmed galaxies from the literature with robust CIII]  equivalent width (W$\rm_{CIII],0}$) constraints. 
We limit the compilation to those systems with known redshifts and spectroscopic constraints that are deep enough to rule out or confirm extreme (i.e., $>$20\AA) equivalent width [CIII],CIII] emission. References: 
[1] \citet{Stark2017}; [2] \citet{Watson2015};[3] \citet{Laporte2017}; [4] \citet{Mainali2017}; [5] \citet{Schmidt2017}; [6] \citet{Stark2015a}; [7] \citet{Stark2015b};  [8] \citet{Ding2017}}
\end{table*}

\section{Discussion and Conclusions}

\subsection{The Origin of NV Emission in EGSY8p7}
 
In section \S3.1, we presented the detection of NV$\lambda$1243 in the spectrum of EGSY8p7, a 
color-selected Ly$\alpha$ emitting galaxy at $z_{\rm{Ly\alpha}}=8.683$ (RB16, \citealt{Zitrin2015}).  
The NV line can have its origin in either HII regions or stellar winds.  The stellar wind feature is 
commonly seen in high redshift galaxy spectra (e.g., \citealt{Shapley2003, Jones2012, 
Steidel2016, Rigby2018}).   Population synthesis models 
including  single stellar populations (e.g., \citealt{Leitherer1999}) and binary stars \citep{Eldridge2016} both 
predict that the NV stellar wind feature can be prominent, particularly at very young ages.  
But the equivalent width of the NV$\lambda$1243 component in these models does not 
reach higher than 2.5~\AA, less than the value we measure in the spectrum of EGSY8p7.  
The emission component of the stellar wind profile is typically observed to be very broad ($>$1000 km s$^{-1}$) 
reflecting the terminal velocity of the winds.  While it is conceivable that broader wings are  too faint to be 
detected in the J-band spectrum, this would imply a total equivalent width that is even larger than 
what we report, putting the observations further at odds with the stellar wind model predictions.    
We thus conclude that the line observed in  EGSY8p7 is most likely to be nebular in origin.   Nevertheless 
the NV detections that are now emerging at $z>7$  motivate both theoretical and observational 
investigation of the range of NV equivalent widths and line widths that can produced by the winds 
of massive stars. 

Powering nebular NV emission  requires an EUV spectrum with substantial flux above 77 eV.
Because the spectra of hot stars have a  strong break above 54 eV, the presence of  
NV in EGSY8p7 likely points to either AGN activity or fast radiative shocks.  The flux ratio of UV 
lines can help to distinguish between the two (e.g., \citealt{Groves2004, Allen2008,Feltre2016}).
The non-detections of CIV and He II in EGSY8p7 allow us to place upper limits on the NV/CIV 
flux ratio ($>$2.6[1.6] at 3[5]$\sigma$) and NV/He II flux ratio ($>$3.8[2.2] at 3[5]$\sigma$), 
provided that both lines are not obscured by skylines.  While this is not a concern for CIV (see Figure 2), 
there is a strong skyline that covers 19\% of the predicted He II spectral window.   
If the NV$\lambda$1243 feature has the same redshift as He II, the latter would indeed 
be located very close to this skyline.  We thus consider the NV/He II limit to be less robust than NV/CIV.  
The constraints on the line ratios in EGSY8p7 are broadly consistent with those in COSY, 
another Ly$\alpha$ emitter from the RB16 sample with NV emission \citep{Laporte2017} and several 
luminous Ly$\alpha$ emitters at $z\sim2-3$ \citep{Sobral2018}. 
Both sources have NV/CIV flux ratios slightly greater than unity, in 
marginal tension with many AGN and shock models with standard abundance ratios 
(e.g., \citealt{Groves2004, Allen2008,Feltre2016,Laporte2017}).   Nevertheless 
there are examples  of AGNs with NV/CIV$\simeq 1$ in high redshift  samples from the 
Sloan Digital Sky Survey \citep{Alexandroff2013}. 
As noted in Laporte et al. (2017), additional physics or resonant scattering of CIV photons might 
ultimately be required to explain the NV/CIV flux ratios that are emerging at $z>7$.  
While distinguishing between shocks and AGN is clearly difficult with only an NV detection, the launch of 
{\it JWST} will provide a wealth of UV and optical spectral constraints on EGSY8p7, clarifying the 
origin of the intense line emission.

With the  NV detection in the \citet{Zitrin2015} $z=8.68$ galaxy, 
three of the thirteen Ly$\alpha$ emitting galaxies known at $z>7$ have been shown to have NV emission 
(e.g.,\citealt{Tilvi2016,Laporte2017}), and two other have recently been detected in Ly$\alpha$ emitters 
at $z=6.9$ \citep{Hu2017} and $z=6.6$ \citep{Sobral2017a}.  The $z>7$ systems with NV emission are among the most luminous 
(1.3-2.5$\times$ greater than L$^\star_{\rm{UV}}$) and massive galaxies known at $z>7$.   It is perhaps not 
surprising that signs of supermassive black hole growth would be detected in such massive early systems.
At lower redshifts AGNs are present in Ly$\alpha$ emitter samples, but the AGN fractions are 
generally less than 5-10\% (e.g., \citealt{Ouchi2008, Zheng2010, Sobral2017, Shibuya2017}). 
But recent work indicates that this fraction increases in the more luminous  Ly$\alpha$ emitters \citep{Sobral2018}. 
While further work is certainly needed to isolate the powering mechanism of the NV detections, these 
early studies appear to indicate that the known Ly$\alpha$ emitters at $z>7$ contain a large 
fraction of sources with hard ionizing spectra, as might be expected if the extreme radiation field 
enhances the visibility of Ly$\alpha$ through the partially neutral IGM.

\subsection{The Frequency of CIII] and CIV Emission at $z>7$}

The UV metal line emission observed in $z>6$ galaxies \citep{Stark2015a, Stark2015b,Stark2017,
Mainali2017, Schmidt2017, Tilvi2016, Laporte2017, Matthee2017} is a marked departure from the 
weak line emission typically found at lower redshift (e.g., \citealt{Shapley2003, Rigby2015, 
Du2017,Du2018}), motivating theoretical investigations into the factors regulating UV line spectra 
\citep{Jaskot2016, Gutkin2016, Nakajima2017,Volonteri2017,Byler2018} and observational efforts to identify intense 
UV line emitters locally and at $z\simeq 1-3$ \citep{Stark2014, Rigby2015, Du2017, Senchyna2017,Maseda2017,LeFevre2017,Amorin2017}.  
With a larger sample of $z>5$ galaxies with UV metal line constraints (see Table 2), we 
can now begin to investigate to what extent the extreme UV line emitters are representative 
among reionization-era sources.  

While CIV emission has been absent in the sources with likely NV detections, the nebular line has 
been detected in two star forming galaxies at $z>6$ \citep{Stark2015b, Mainali2017, Schmidt2017}.  
In contrast to the galaxies with NV emission, the CIV emitters are among the lowest luminosity galaxies with 
detectable Ly$\alpha$ emission  in the reionization era.   Both are gravitationally lensed with 
absolute UV magnitudes (M$_{\rm{UV}}$=$-20.1$, $-19.3$) that are 2.8-5.8$\times$ lower than the 
median luminosity of the three $z>7$ sources with reported NV emission.  The presence of CIV 
requires a significant supply of 48 eV photons capable of triply-ionizing carbon. 
Local UV spectral databases  (e.g., \citealt{Berg2016, Senchyna2017}) show that nebular CIV emission 
becomes commonplace in actively star forming galaxies with gas-phase metallicities below 
12+log O/H $\simeq 8.0$, a consequence of the 
rapid transition in the hardness of the EUV spectrum of massive stars below this metallicity threshold \citep{Senchyna2017}.
The sharp increase in hard photons can be explained in part by less dense stellar winds (which allow 
more of the EUV flux to escape the stellar atmospheres) and more efficient evolutionary pathways toward 
the creation of hot stripped stars at low metallicity (e.g., \citealt{Senchyna2017}). 
The detection of strong nebular CIV emission may thus be able to provide a valuable signpost of low metallicity 
stellar populations.   In Figure 6, we show the relationship between 
M$_{\rm{UV}}$ and CIV equivalent width for galaxies at $z\gtrsim 5$, updating a plot originally  
presented in \citet{Shibuya2017} with the new upper limits presented in this paper.  The presence of 
CIV is limited to the lowest luminosity systems, as might be expected if a luminosity-metallicity 
relationship is present at these early times.  In this context, the absence of strong 
CIV emission in the luminous  RB16 sources may  indicate that metallicity of massive stars in 
these systems has already been polluted above the threshold necessary for powering a 
hard EUV spectrum.  {\it JWST} will soon begin to provide constraints on the rest-optical lines 
in these and other reionization-era galaxies, making it possible to investigate the relationship 
between high ionization UV lines and metallicity in more detail.

\begin{figure}
\centering
\begin{tabular}{c}

\hspace{-0.6 cm}

\subfloat{ \includegraphics[angle=90,scale=0.39]{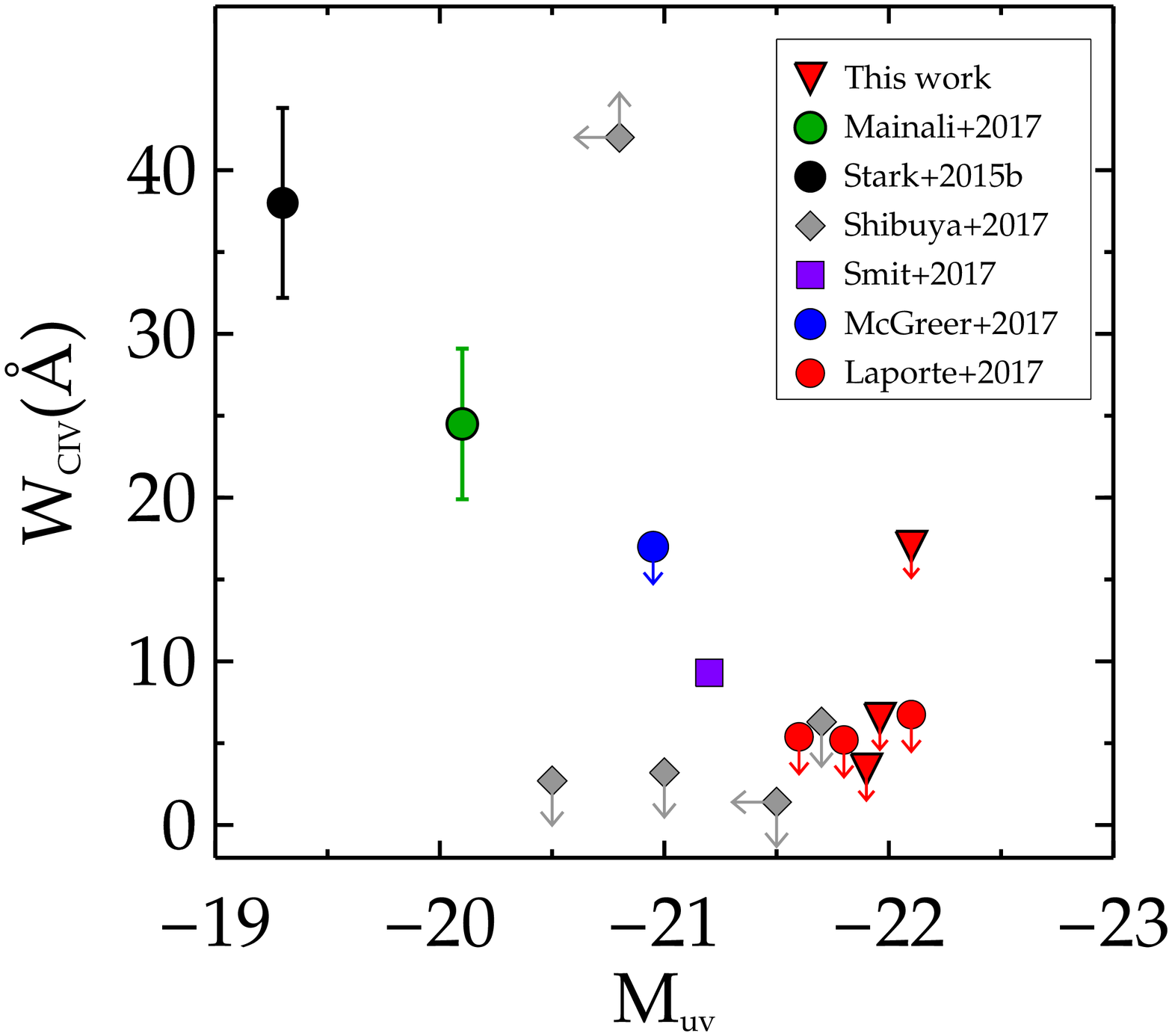}}

\end{tabular}
\caption{ Plot of rest frame CIV equivalent width (W$\rm_{CIV}$) as a function of M$\rm_{UV}$. The red triangles are $z\sim7-9$ sources from this paper. 
The other symbols are compilation from the literature: red circle \citet{Laporte2017}; grey diamond \citet{Shibuya2017}; violet square \citet{Smit2017}; blue circle \citet{McGreer2017}; green circle \citet{Mainali2017}; black circle \citet{Stark2015b}. These existing studies reveal CIV emission only in the lowest 
luminosity galaxies, consistent with trends found locally and at lower redshifts.   }

\end{figure}

\begin{figure}
\centering
\begin{tabular}{c}

\hspace{-0.6 cm}
\subfloat{ \includegraphics[angle=90,scale=0.39]{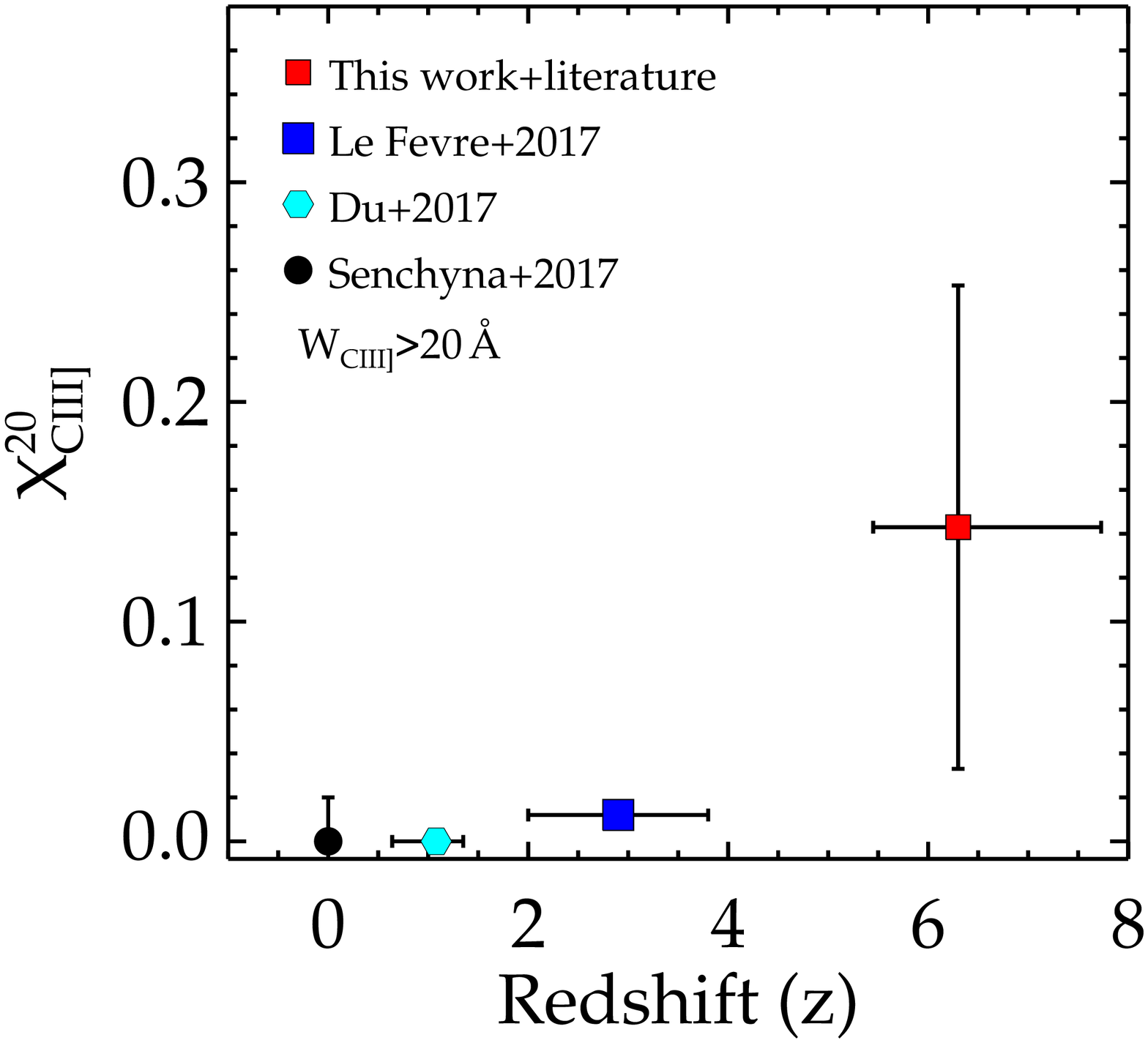}}

\end{tabular}
\caption{ Fraction of strong CIII] emitters (W$\rm_{CIII]}$>20~\AA) as a function of redshift. The fraction at $z\sim0$, $z\sim1$ and $z\sim3$ are computed using data from \citet{Senchyna2017}, \citet{Du2017} and \citet{LeFevre2017}, respectively.
The fraction at $z\sim5.5-9$ is calculated using data in Table 2 and Table 3. We note that the parent sample at $z\sim5.5-9$ is derived from spectroscopically confirmed sources with constraints on CIII] emission. }

\end{figure}

Much of the attention on metal lines has focused on the [CIII], CIII] doublet, as it is often the strongest line 
in the UV other than Ly$\alpha$.   Intense [CIII], CIII] emission (EW=22\AA) has been 
detected in EGS-zs8-1, the $z=7.73$ Ly$\alpha$ emitter from the RB16 sample 
\citep{Oesch2015, Stark2017}.  Similarly strong CIII] (EW=22.5~\AA) is present in A383-5.2 \citep{Stark2015a}, a gravitationally lensed $z=6.03$ galaxy with Ly$\alpha$ \citep{Richard2011}.  In this paper, we 
present upper limits on the strength of CIII] in five sources at $z>5.4$.  
When taken  together with robust limits from the literature (see Table 3), 
we find that the fraction of galaxies   at $z>5.4$ with CIII] EW$>20$~\AA\ is $0.14\pm0.10$.  
While such strong CIII] emission is clearly not ubiquitous among reionization-era galaxies, it is 
even less common among systems at lower redshifts.  In a study of 2543 galaxies at 
$2<z<3.8$, \citet{LeFevre2017} find only 31 sources with CIII] EW in excess of 20~\AA, 
implying a fraction of just 1.2\%.  Similarly low fractions are found in other studies of galaxies 
at $z\simeq 0-3$ \citep{Rigby2015, Du2017, Senchyna2017}.  
These preliminary indications of evolution in the fraction of extreme EW CIII] emitters (Figure 7) 
could be interpreted as a byproduct of the changing demographics of star forming galaxies; as  
larger sSFR and more intense optical line emission become more 
common at early times, strong CIII] emission should similarly become more typical (e.g., \citealt{Senchyna2017,
Maseda2017}).   Others have put forth an alternative explanation, arguing  
that sources with CIII] EW$>20$~\AA\ likely have AGN activity \citep{Nakajima2017}; in this picture 
the increased incidence of extreme CIII] emitters may reflect the changing nature of Ly$\alpha$ 
emitters at $z\simeq 7-8$. One other important factor that could possibly govern the strength of carbon lines is the relative C/O abundance \citep{Stark2014,Berg2016,Nakajima2017,LeFevre2017,Berg2018}. 
These papers show that C/O decreases at lower O/H over 0.1$\leq$Z/Z$_{\odot}$$\leq$1.0. If C/O is very low at $z>6$, it could cause the lines to be weaker than expected. However, measurements of C/O at  $z>6$ are 
not yet possible with current facilities. Future observations that simultaneously detect CIII], CIV and OIII] at $z>6$ will shed light on the 
role of C/O abundance in shaping the strength of carbon lines in these early galaxies.

While our knowledge of the UV line statistics is undoubtedly still in its infancy, the presence of 
intense  line emission is beginning to refine our understanding of the nature of the 
Ly$\alpha$ emitter population in the reionization era.   Prominent emission lines (NV, CIV, or CIII]) 
have now been reported in five of the thirteen Ly$\alpha$ emitters known at $z>7$.   Galaxies with 
similar Ly$\alpha$ EW at lower redshifts (e.g., \citealt{Shapley2003, Jones2012, Du2017, 
LeFevre2017}) very rarely exhibit such extreme UV line emission.  These results continue to suggest 
that the radiation field is playing an important role in making these systems visible in 
Ly$\alpha$ emission in an epoch where most Ly$\alpha$ photons are strongly attenuated by the IGM.  
Attempts to map the evolving Ly$\alpha$ EW distribution to a neutral hydrogen fraction must ultimately 
take into account these variations in the radiation field.   The launch of {\it JWST} will soon provide an 
opportunity to extend rest-UV studies to larger samples, increasing our understanding of the factors 
regulating the escape of Ly$\alpha$ in the reionization era.   While many of the spectroscopic studies  
will focus on the strong rest-optical lines, the UV should not be neglected as the presence of various 
high ionization features (NV, He II, CIV) provides the most direct path toward detecting extreme 
sources, whether they be AGN or very low metallicity stellar populations.

\section*{Acknowledgements}
We thank referee for the useful suggestions and comments. We are very grateful to Anna Feltre for enlightening conversations.
DPS acknowledges support from the National Science Foundation  through the grant AST-1410155.  RSE and NL acknowledge
support from the European Research Council through an Advanced Grant FP7/669253.
JR acknowledges support from the European Research Council through a Starting grant FP7/336736.
This work was partially supported  by a NASA Keck PI Data Award, administered by the 
NASA Exoplanet Science Institute. Data presented herein were obtained at the W. M. Keck Observatory 
from telescope time allocated to the National Aeronautics and Space Administration through the agency's 
scientific partnership with the California Institute of Technology and the University of California. 
The Observatory was made possible by the generous financial support of the W. M. Keck Foundation.
The authors acknowledge the very significant cultural role that the
summit of Mauna Kea has always had within the indigenous Hawaiian community.
We are most fortunate to have the opportunity to conduct observations from this mountain. 
Also based on observations made with ESO Telescopes at the La Silla Paranal Observatory under programme ID 092.A-0630(A).
The LBT is an international collaboration among institutions in the United States, Italy and Germany. 
LBT Corporation partners are: The University of Arizona on behalf of the Arizona university system; Istituto 
Nazionale di Astrofisica, Italy; LBT Beteiligungsgesellschaft, Germany, representing the Max-Planck Society, 
the Astrophysical Institute Potsdam, and Heidelberg University; The Ohio State University, and The Research 
Corporation, on behalf of The University of Notre Dame, University of Minnesota and University of Virginia.

\bibliographystyle{mnras}
\bibliography{references}

\appendix

\label{lastpage}
\end{document}